\documentclass[mathleft]{an}
\usepackage{times}
\usepackage{graphicx}
\usepackage{url}
\usepackage{bm}
\graphicspath{{./fig/}{./png/}}

\newcommand{\EQ}{\begin{equation}}
\newcommand{\EN}{\end{equation}}
\newcommand{\EQA}{\begin{eqnarray}}
\newcommand{\ENA}{\end{eqnarray}}
\newcommand{\eq}[1]{(\ref{#1})}

\newcommand{\Eq}[1]{Eq.~(\ref{#1})}
\newcommand{\Eqs}[2]{Eqs.~(\ref{#1}) and~(\ref{#2})}

\newcommand{\eqss}[2]{(\ref{#1})--(\ref{#2})}

\newcommand{\App}[1]{Appendix~\ref{#1}}
\newcommand{\Sec}[1]{Sect.~\ref{#1}}

\newcommand{\Fig}[1]{Fig.~\ref{#1}}

\newcommand{\Tab}[1]{Table~\ref{#1}}

\newcommand{\meanEMFT}{\overline{\vec{\cal E}}^{\rm T}}
\newcommand{\meanEMF}{\overline{\vec{\cal E}}}
\newcommand{\meanemf}{\overline{\cal E}}
\newcommand{\hatemf}{\hat{\cal E}}
\newcommand{\hatB}{\hat{B}}
\newcommand{\hatBB}{\hat{\bm{B}}}

\newcommand{\meanA}{\overline{A}}
\newcommand{\meanB}{\overline{B}}

\newcommand{\meanAA}{\overline{\vec{A}}}
\newcommand{\meanBB}{\overline{\vec{B}}}
\newcommand{\meanBBT}{{\overline{\vec{B}}}^{\rm T}}
\newcommand{\meanJJ}{\overline{\vec{J}}}
\newcommand{\meanUU}{\overline{\vec{U}}}

\newcommand{\uu}{{\bm{u}}}

\newcommand{\bb}{{\bm{b}}}
\newcommand{\aaT}{{\bm{a}}^{\rm T}}
\newcommand{\bbT}{{\bm{b}}^{\rm T}}

\newcommand{\ee}{{\bm{e}}}
\newcommand{\qq}{{\bm{q}}}

\newcommand{\kk}{{\bm{k}}}

\newcommand{\xx}{{\bm{x}}}

\newcommand{\nab}{\mbox{\boldmath $\nabla$} {}}

\newcommand{\tensor}[1]{\boldsymbol{\mathsf #1}}
\newcommand{\aTens}{\tensor{\alpha}}
\newcommand{\eTens}{\tensor{\eta}}

\newcommand{\MMMM}{\mbox{\boldmath ${\sf M}$} {}}

\newcommand{\ii}{{\rm i}}

\newcommand{\dd}{{\rm d} {}}
\newcommand{\const}{{\rm const}  {}}

\def\Rm{R_{\rm m}}

\def\cs{c_{\rm s}}
\def\kf{k_{\rm f}}
\def\urms{u_{\rm rms}}

\def\etatz{\eta_{\rm t0}}
\def\etat{\eta_{\rm t}}

\def\tauto{\tau_0}
\def\Iota{\xi}
\def\Mu{\eta_{\scriptscriptstyle\cal E}}  

\def\half{{\textstyle{1\over2}}}

\newcommand{\Mm}{\,{\rm Mm}}

\newcommand{\yapj}[3]{: #1, {ApJ} {#2}, #3}

\newcommand{\yan}[3]{: #1, {AN} {#2}, #3}

\newcommand{\yana}[3]{: #1, {A\&A} {#2}, #3}

\newcommand{\ygafd}[3]{: #1, {GApFD} {#2}, #3}

\newcommand{\ypf}[3]{: #1, {PhFl} {#2}, #3}

\newcommand{\yprl}[3]{: #1, {PhRvL} {#2}, #3}
\newcommand{\ypre}[3]{: #1, {PhRvE} {#2}, #3}

\newcommand{\ymn}[3]{: #1, {MNRAS} {#2}, #3}

\newcommand{\yjour}[4]{: #1, {#2} {#3}, #4}

\newcommand{\ybook}[3]{: #1, {\it #2} (#3)}

\Pagespan{725}{}
\Yearpublication{2011}
\Yearsubmission{2010}
\Month{1}
\Volume{332}
\Issue{1}
\DOI{10.1002/asna.200811027}

\begin{document}

\title{Modeling spatio-temporal nonlocality in mean-field dynamos}
\authorrunning{M. Rheinhardt \& A. Brandenburg}
\author{
M. Rheinhardt$^{1,2}$, and
A. Brandenburg\thanks{Corresponding author: brandenb@nordita.org}$^{1,3}$,
}
\institute{
$^1$Nordita\thanks{Nordita is a Nordic research institute jointly operated
by the Stockholm University and the Royal Institute of Technology, Stockholm.},
    AlbaNova University Center, Roslagstullsbacken 23,
    SE-10691 Stockholm, Sweden\\
$^2$Department of Physics, Gustaf H\"allstr\"omin katu 2a (PO Box 64),
    FI-00014 University of Helsinki, Finland\\
$^3$Department of Astronomy, Stockholm University,
    SE-10691 Stockholm, Sweden
}

\received{2011 Sep 29}  \accepted{2011 Oct 18}
\publonline{2011 Dec 30}

\keywords{magnetic fields -- magnetohydrodynamics (MHD)} 

\abstract{%
When scale separation in space and time is poor, the $\alpha$ effect
and turbulent diffusivity have to be replaced by integral kernels.
Earlier work in computing these kernels using the test-field method
is now generalized to the case in which both spatial and temporal
scale separations are poor.
The approximate form of the kernel is such that it can be treated
in a straightforward manner by solving a partial differential equation
for the mean electromotive force.
The resulting mean-field equations are solved for oscillatory $\alpha$--shear
dynamos as well as $\alpha^2$ dynamos in which $\alpha$ is antisymmetric
about the equator, making this dynamo also oscillatory.
In both cases, the critical values of the dynamo number is lowered
by the fact that the dynamo is oscillatory.
\keywords{MHD -- Turbulence}}

\maketitle

\section{Introduction}

Mean-field dynamo theory describes the evolution of the averaged
magnetic field.
This theory is relevant for the understanding of the origin of ordered
magnetic fields in the Sun and other late-type stars.
Compared to the original induction
equation, the averaged equation contains extra
terms which capture the effects of systematic correlations between
velocity and magnetic field fluctuations.
Some of these terms (for example the $\alpha$ effect) can be
responsible for the generation of mean magnetic fields.

Mean-field dynamo theory provides an important tool for a number of
astrophysical applications.
However, it also suffers from several shortcomings, some of which
can be the result of simplifications that are not
well justified and often not even necessary.
In this paper we focus on the issue of poor scale separation in
space and time.
Broadly speaking, if there is poor scale separation, multiplications
with mean-field coefficients must be replaced by convolutions
with corresponding integral kernels.
Obviously, as far as temporal scale separation is concerned, this
effect cannot be very important for the Sun, because the cycle time
is much longer than the convective turnover time.
However, with respect to
spatial scale separation this is no longer true, because
at the bottom of the solar convection zone the pressure scale height
and with it the typical size of the convection cells is $50\Mm$,
and hence comparable to the depth of the convection zone of $200\Mm$
which is also the scale of the mean magnetic field.
Although the concept of writing the mean electromotive force as a
spatio-temporal convolution with the mean magnetic field was well known
(e.g., R\"adler 1976), there was the problem that, until recently, not
much was known about the form of the integral kernels that are to be used.
In the past there have been several attempts to compute the integral
kernels from turbulence simulations (e.g., Miesch et al.\ 2000;
Brandenburg \& Sokoloff 2002), but the situation has changed drastically
with the advent of the test-field method (Schrinner et al.\ 2005, 2007)
which allowed an accurate determination of the integral kernels in
space (Brandenburg et al.\ 2008) and time (Hubbard \& Brandenburg 2009).
As a result, we now know that the kernels of most of the components of
the $\aTens$ and $\eTens$ tensors are Lorentzians in spectral space and
exponentials in real space (Brandenburg et al.\ 2008).
It turns out that in these simple cases,
the resulting integro-differential equation for the magnetic field
can be reformulated into a set of two coupled differential equations
of parabolic type, one for the magnetic
field and one for the electromotive force.

In hindsight, we can say that even temporal scale separation can sometimes
be relevant, because nowadays we are not only comparing with the Sun and
other astrophysical bodies, but also with direct numerical simulations (DNS).
In DNS we may well have situations in which the dynamo $e$-folding times
and perhaps also the cycle periods become comparable to the turnover time
of the turbulence.
In these more extreme situations, we have much better possibilities of
testing theory.
Furthermore, with DNS there is more freedom in constructing cases that
may be hard to find in real astrophysical bodies, but for which the same
mean-field theory should equally well be applicable.
Furthermore, DNS allow us to determine turbulent transport
coefficients to high accuracy, facilitating therefore detailed comparison
with mean-field theory.
Indeed, it turns out that in DNS the growth rates of dynamos
can well be comparable to the turnover time.
A dramatic example was presented by Hubbard \& Brandenburg (2009),
where the growth rate of a Roberts flow dynamo was found to be
significantly different from the value expected based on the
analytic dispersion relation using coefficients that have been
determined numerically using the test-field method,
but under the assumption of perfect scale separation in time.

\section{Formalism}

To set the scene, let us begin with the mean-field dynamo equation
for the mean magnetic field $\meanBB$,
\EQ
{\partial\meanBB\over\partial t}=\nab\times\left(
\meanUU\times\meanBB+\meanEMF-\eta\mu_0\meanJJ\right),
\label{dynamoeq}
\EN
where $\meanUU$ is the mean velocity,
$\meanEMF$ is the mean electromotive force,
and $\meanJJ=\nab\times\meanBB/\mu_0$ is the mean current density,
with $\mu_0$ being the vacuum permeability,
and $\eta$ the microscopic (molecular) magnetic diffusivity.
Under certain conditions, $\meanEMF$ can be expanded in terms of
the mean magnetic field and its derivatives as
\EQ
\meanemf_i=\alpha_{ij}\meanB_j+\eta_{ijk}\meanB_{j,k}+...,
\label{emf_expansion}
\EN
where the comma denotes partial differentiation and the
dots refer to higher spatial derivatives of $\meanBB$,
temporal derivatives of $\meanBB$, as well as terms
independent of $\meanBB$.

In many cases of practical interest, only the lowest (including the zeroth)
order spatial derivatives are retained, because they are sufficient
for capturing qualitatively new effects such as large-scale dynamo action.
This has led to a large number of mean-field dynamo models
that were applied to the Sun, other stars, galaxies, and
even accretion discs.
In such models, the length scales of the resulting mean field
become often quite small, especially in the nonlinear regime;
see, e.g., Chatterjee et al.\ (2011a, Figs.~9--11).
In this context, `small' means that the scale of the mean field
becomes comparable to and even smaller than
the scale of the energy-carrying eddies.
In stratified turbulence, as present in the Sun,
the scale of these eddies is
often assumed to be proportional to the local pressure scale height,
which is about $50\Mm$ at the bottom of the solar convection zone.
However, in mean-field models the magnetic fields show frequently
variations on scales much smaller than this.
Chatterjee et al.\ (2011a), discussed the small-scale
fields at the bottom of the convection zone in their simulations
of a mean-field dynamo model
as an artifact of the neglect of nonlocality in space,
but no solution to this problem was feasible at the time.

Looking at \Eq{emf_expansion}, it is clear that higher spatial derivatives
need to be retained when the mean field is no longer slowly varying in space.
Unfortunately, such a series expansion becomes easily quite cumbersome, and
it is then better to replace \Eq{emf_expansion} by a convolution of the
mean magnetic field $\meanBB$ with some integral kernel.
As alluded to above, a representation of $\meanEMF$ in terms of a
convolution of $\meanBB$ with a kernel
determined by the statistical properties of the turbulence
has long been known to be the more basic one (e.g., R\"adler 1976).
By allowing the convolution to be also over time, we can automatically
include all temporal derivatives as well, i.e., we can instead
of \Eq{emf_expansion} write
\EQ
\meanemf_i(\xx,t)=\int G_{ij}(\xx,\xx',t,t')\meanB_j(\xx',t')
\,\dd^3x'\,\dd t',
\EN
where we have again ignored terms that are independent of $\meanBB$.

For simplicity, we shall restrict ourselves now to statistically
homogeneous and steady turbulence, in which case $G_{ij}$ is
translation invariant in space and time and depends thus only
on the arguments $\xx-\xx'$ and $t-t'$.
In cases with boundaries, this is not possible, but the formalism
presented below can easily be adapted to such cases as well; see
Chatterjee et al.\ (2011b).

Continuing now with the translation invariant case,
the convolution becomes a multiplication in Fourier space, i.e.,
\EQ
\hatemf_i(\kk,\omega)=\hat{G}_{ij}(\kk,\omega)\hatB_j(\kk,\omega),
\EN
where hats indicate Fourier transformation in space and time, e.g.,
\EQ
\hatemf_i(\kk,\omega)=\int\meanemf_i(\xx,t) e^{\ii(\kk\cdot\xx-\omega t)}
\,\dd^3 x\,\dd t.
\EN
In view of the traditional distinction of contributions to $\meanEMF$ from
the $\alpha$ effect and turbulent diffusivity, it is convenient to
write the Fourier transform of the kernel in the form
\EQ
\hat{G}_{ij}(\kk,\omega)
=\frac{\alpha^{(0)}_{ij}+\eta^{(0)}_{ijk}\ii k_k}
{\hat{D}(\kk,\omega)},
\label{Gij}
\EN
where $\alpha^{(0)}_{ij}$ and $\eta^{(0)}_{ijk}$ are assumed
to be tensors that are independent of $\kk$ and $\omega$.
The goal of this paper is to verify the approximate validity of \Eq{Gij}
and to consider the consequences of such a structure
for mean-field dynamo models.

In the following we consider triply periodic domains and define
mean fields as planar averages over the $x$ and $y$ directions,
so that $\meanBB$ is only a function of $z$ and $t$.
In that case, $\kk=(0,0,k)$ has only one component.
Recent work of Hubbard \& Brandenburg (2009) has already revealed that
for fixed $k$, $\hat{D}(k,\omega)$ is proportional to $1-\ii\omega\tau$,
where $\tau$ is a fit parameter that is approximately equal to
the turnover time, i.e., $\tau\urms\kf\approx1$, where $\urms$
is the rms velocity of the turbulence and $\kf$
is the wavenumber of its energy-carrying eddies
On the other hand, for $\omega=0$, $\hat{D}(k,\omega)$ is
approximately proportional to $1+(ak/\kf)^2$, where $a$ is
a dimensionless parameter, for which values between 0.2 and 1
have been found over a range of different simulations
(Brandenburg et al.\ 2008, 2009; Madarassy \& Brandenburg 2010).
Consequently,
we propose in the present paper that $\hat{D}(k,\omega)$ can be
approximated by
\EQ
{\hat{D}(k,\omega)}=
1-\ii\omega\tau+\ell^2 k^2 .
\label{Dko}
\EN
with the additional parameter $\ell$ having the dimension of a length.
Such a form, even if it is still an approximation that neglects
higher powers of $k$ and $\omega$, has already the advantage of
alleviating problems of unrealistic variations of the magnetic field
on short length and time scales.
Moreover, it leads to an easily treatable
partial differential equation for $\meanEMF$ in real space, namely
\EQ
\left(1+\tau{\partial\over\partial t}
-\ell^2{\partial^2\over\partial z^2}\right)\meanemf_i
=\alpha^{(0)}_{ij}\meanB_j+\eta^{(0)}_{ijk}\meanB_{j,k}.
\label{emf_nonlocal}
\EN
Note that in the limit $\tau\to0$ and $\ell\to0$, the usual dynamo
equations are recovered.
Thus, nonlocality is captured simply by specifying $\tau$ and $\ell$,
while the tensors $\alpha^{(0)}_{ij}$ and $\eta^{(0)}_{ijk}$ can simply be
regarded as the usual ones of $\alpha$ effect and turbulent diffusivity
for the limit $k\rightarrow 0$, $\omega \rightarrow 0$.
Therefore, the superscripts $(0)$ will from now on be dropped.
The purpose of this paper is to establish not only the validity
of this approach, but also to assess the properties of 
mean-field dynamos when $\meanEMF$ is obtained as the solution of the
evolution equation \Eq{emf_nonlocal}.

\section{The kernel function $\hat{D}(k,\omega)$ from DNS}

\subsection{Turbulence in a periodic domain}
\label{PeriodicDomain}

In the following we present results for three-dimensional
isothermal turbulence that is being forced in a narrow range of
wavenumbers around a representative wavenumber $\kf$.
We adopt a cubic domain of size $L^3$,
measure length in units of the inverse minimal box wavenumber
$k_1=2\pi/L$ and choose $\kf/k_1\approx2.2$.
We vary the magnetic Reynolds number,
\EQ
\Rm=\urms/\eta\kf,
\EN
where $\urms$ is the rms velocity of the turbulence,
keeping the rms Mach number, $\urms/\cs$ at around 0.1.
In agreement with the considerations above,
time is expressed in units of the turnover time,
defined here as $\tauto=(\urms\kf)^{-1}$, and the turbulent
magnetic diffusivity is expressed in units of $\etatz=\urms/3\kf$
(cf.\ Sur et al.\ 2008).

\subsection{Test-fields in space and time}

To establish the form of \Eq{Dko} we use the test-field method,
i.e., we solve, for a given turbulent velocity field,
the equations governing the departure of the magnetic field
from a given mean field,
that is, we determine the magnetic {\it fluctuations} $\bb$ caused
by the interaction of the turbulent velocity with the mean field.
This mean field is referred to as the test field
and is marked by the superscript T.
For each test field $\meanBBT$, we find a corresponding departure
$\bbT=\nab\times\aaT$ by solving the inhomogeneous equation
for the corresponding vector potential $\aaT$,
\EQ
{\partial\aaT\over\partial t}=
\meanUU\times\bbT+\uu\times\meanBBT+\left(\uu\times\bbT\right)'
+\eta\nabla^2\aaT,
\EN
where $\left(\uu\times\bbT\right)'=\uu\times\bbT-\overline{\uu\times\bbT}$ is the
fluctuating part of $\uu\times\bbT$, and compute the corresponding
mean electromotive force, $\meanEMFT=\overline{\uu\times\bbT}$.
We use test fields that are harmonic functions
with wavenumber $k$ and frequency $\omega$ and
point either in the $x$ or in the $y$ direction, i.e.,
\EQ
\meanBB^{i{\rm c}k\omega}\!\!=\ee_i\cos kz\cos\omega t,\quad  
\meanBB^{i{\rm s}k\omega}\!\!=\ee_i\sin kz\cos\omega t,
\EN
$i=1,2$, where $\ee_1$ and $\ee_2$ are unit vectors pointing in the
$x$ and $y$ directions, respectively.
The third component is here without interest, because
$\nab\cdot\meanBB=\partial\meanB_z/\partial z=0$, so
$\meanB_z=\const$, and is chosen to be zero initially.

\begin{figure}[t!]\begin{center}
\includegraphics[width=\columnwidth]{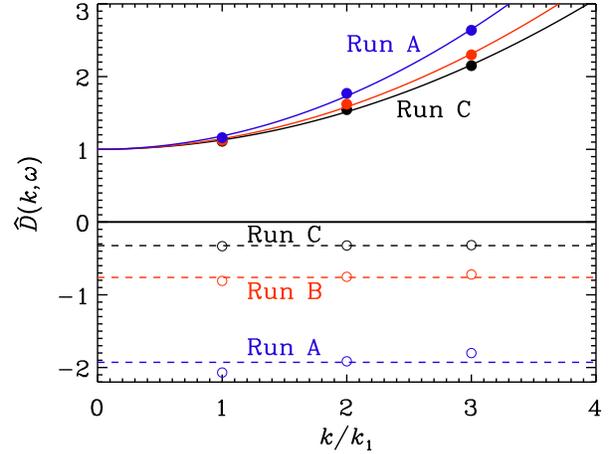}
\end{center}\caption[]{ 
$\hat{D}(k,\omega)$ for $\omega\tauto=1.04$ (Run~A),
0.52 (Run~B), and 0.26 (Run~C).
Open and filled circles denote the real and imaginary parts of
$\hat{D}(k,\omega)$ as obtained from the test-field method;
the parabolas give a fit proportional to $1+\ell^2 k^2$.
Dashed lines: average of the three data points
of the imaginary part of $\hat{D}(k,\omega)$ for each $\Rm$.
For the fit parameters $\ell$ and $\tau$ see \Tab{Simulations}.
}\label{psummary}\end{figure}

\begin{table}[b!]\caption{
Summary of fit parameters for runs without helicity (Runs~A--D)
and one with helicity (Run~E) using either \Eq{foreta} or \Eq{foralp}.
}\vspace{12pt}\centerline{\begin{tabular}{c|ccccccc}
Run & $\Rm$ & $\omega\tauto$ & $\tau/\tauto$ & $\ell\kf$ & Equation \\
\hline
\hline\\[-2.5mm]
 A  &  8   &      1.04      &      1.85     &    0.99   & \eq{foreta} \\ 
 B  &  8   &      0.52      &      1.46     &    0.88   & \eq{foreta} \\ 
 C  &  8   &      0.26      &      1.24     &    0.83   & \eq{foreta} \\ 
 D  & 53   &      0.38      &      1.21     &    0.77   & \eq{foreta} \\ 
 E  & 57   &      0.35      &      0.67     &    0.60   & \eq{foreta} \\ 
 E  & 57   &      0.35      &      0.59     &    0.80   & \eq{foralp} 
\label{Simulations}\end{tabular}}\end{table}

Using the standard test-field method, we obtain directly the tensors
$\hat\alpha_{ij}(k,\omega)$ and $\hat\eta_{ijk}(k,\omega)$.
From that we can determine $\hat{D}$ for different values of
$k$ and $\omega$ according to 
\EQ
\hat{D}(k,\omega) = \hat\alpha_{ij}(0,0)/\hat\alpha_{ij}(k,\omega),
\label{foralp}
\EN
or
\EQ
\hat{D}(k,\omega) = \hat\eta_{ij}(0,0)/\hat\eta_{ij}(k,\omega),
\label{foreta}
\EN
employing the known values $\hat\alpha_{ij}(0,0)$ or $\hat\eta_{ij}(0,0)$.
Furthermore, since we consider isotropic turbulence, both tensors
are isotropic, i.e., $\hat\alpha_{ij}=\hat\alpha\delta_{ij}$ and
$\hat\eta_{ijk}=\hat\etat\epsilon_{ijk}$, but $\hat\alpha=0$
for non-helical turbulence (Runs~A--D of \Tab{Simulations}).
For this case $\hat{D}(k,\omega)$ is shown in \Fig{psummary},
where we plot its real and imaginary parts for the scale separation ratio
$\kf/k_1=2.2$, $\Rm=8$, and three values of $\omega\tauto$.
The real part of ${\hat{D}(k,\omega)}$ is a fit to a profile of the form
$1+\ell^2 k^2$,
while the imaginary part of ${\hat{D}(k,\omega)}$ is
approximately independent of $k$.
This is consistent with $\Im\{{\hat{D}(k,\omega)}\}=-\omega\tau$,
where $\tau$ is obtained by taking the average
value of $\omega\tau$ for all three $k$ values.
We find that $\tau/\tauto$ and $\ell\kf$ are of the order of unity.
In agreement with the ansatz (\ref{Dko}) the parameter $\ell\kf$ varies
only weakly with $\omega$, but $\tau/\tauto$ shows a stronger variance,
indicating the presence of higher powers of $\omega$ in $\hat{D}$.
Both parameters vary somewhat with $\Rm$; see \Tab{Simulations} for details.
The additional Run~E
differs from Run~D only in including helicity in the forcing and hence in the flow.
For both $\tau$ and $\ell$ the resulting values obtained by using \Eq{foralp} and \Eq{foreta} are similar.
The value of $\ell$ is also similar to that of Run~D
whereas $\tau$ is reduced by a factor of $\approx 2$.

Comparing the results for Runs~D and E suggests that in \Eq{Dko} the values of
$\tau$ are reduced by a factor of 2 when there is helicity in the turbulence,
while $\ell$ remains approximately unchanged.

Thus, in conclusion, we have verified that,
for a turbulent flow such as that considered here,
the integral kernel in \Eq{Gij} with ${\hat{D}(k,\omega)}$ is
roughly given by \Eq{Dko}.
We concede, however, that the modeling of the $\omega$ dependence of $\hat{G}_{ij}$
is worth to be improved taking into account higher orders in $\omega$.
In the remainder of this paper
we examine properties of the resulting mean-field equations.

\section{Application to mean-field dynamo models}

\subsection{Nonlocality in dynamo waves}
\label{Nonlocality}

Some limited insight into the effects of nonlocality for dynamo
waves has already been provided in the paper by Brandenburg et al.\ (2008),
who considered nonlocality in space, but not in time.
Based on their test-field results, they found a kernel compatible
with a Lorentzian in $k$ space.
Generally speaking, such a kernel makes the resulting mean electromotive force
smoother by acting preferentially on the largest scale in the domain.
In the present paper we repeat a similar experiment, but with the
difference that we include here also nonlocality in time.

Nonlocality in time can lead to somewhat unexpected behavior
of oscillatory dynamos of $\alpha$--shear type in that it
enhances their growth rate
and, more importantly, it lowers the critical value for dynamo action.
This is different from the $\alpha^2$ dynamo case, where the
presence of an extra time derivative always leads to a lower
growth rate (Brandenburg et al.\ 2008).
This can be seen by comparing the two dispersion relations for
$\alpha^2$ and $\alpha$--shear dynamos with constant $\alpha$ and shear.
By making an ansatz of the form
\EQ
\meanBB=\hatBB\exp\left[\ii(kz-\omega t)+\lambda t\right], \label{Bansatz}
\EN
with real coefficients $k$ (wavenumber), $\omega$ (frequency),
and $\lambda$ (growth rate), we can easily obtain the dispersion relation
for the system of \Eqs{dynamoeq}{emf_nonlocal} in implicit form.
In the case of an $\alpha^2$ dynamo with $\eta=0$
we have (see \App{Disper})
\EQ
\lambda=\Iota^{-1}\left(\pm|\alpha k|-\etat k^2\right),\quad\omega=0,
\label{lama2}
\EN
where we have introduced the correction factor
\EQ
\Iota=1+\tau\lambda+\ell^2 k^2,
\EN
In the case of a pure $\alpha$--shear dynamo, with the
$\alpha\meanB_x$ term neglected in favor of $S\meanB_x$,
and again $\eta=0$, it is convenient
to seek marginally excited oscillatory solutions with $\lambda=0$,
which gives, with $\Iota=1+\ell^2 k^2$,
\EQ
\omega^2=\half\tau^{-2}\Iota^2\left[-1
+\sqrt{1+(2\tau\etat k^2/\Iota^2)^2}\right],
\label{omaS}
\EN
which allows us to compute the critical dynamo number
\EQ
D_{\rm crit}\equiv\alpha S/\etat^2 k^3
=2\omega(1-\omega^2\tau/\etat k^2)/\etat k^2.
\label{Dcrit}
\EN
Note that for an $\alpha^2$ dynamo the threshold remains unchanged,
while for an $\alpha$--shear dynamo the term $\omega^2\tau$ always
lowers the threshold.

\begin{figure}[t!]\begin{center}
\includegraphics[width=.93\columnwidth]{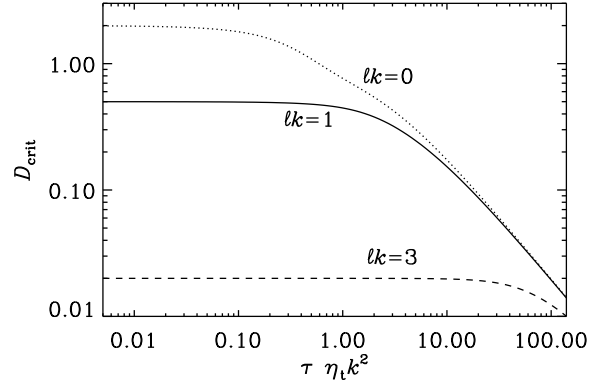}
\end{center}\caption[]{ 
Critical dynamo number for an $\alpha$--shear dynamo as a function
of $\tau\etat k^2$ for different values of $\ell k$.
Microscopic magnetic diffusivity, $\eta$, is here neglected.
}\label{piter_ao2}\end{figure}

In \Fig{piter_ao2} we show $D_{\rm crit}$ for an
$\alpha$--shear dynamo as a function of $\tau\etat k^2$
for different parameters $\ell k$.
For small values of $\tau$, the usual value of $D_{\rm crit}=2$
is obtained (e.g., Brandenburg \& Subramanian 2005).

\subsection{Nonlocality and boundaries}
\label{NonlocalityBound}

We have mentioned in the beginning that the effect of spatial
nonlocality should consist in a spatial smoothing
of the mean electromotive force.
However, the solutions presented so far are all entirely harmonic.
To see the anticipated smoothing effect, we can either
consider nonlinear solutions (as done in Brandenburg et al.\ 2008),
or we can consider solutions with boundaries, which breaks the
monochromatic nature of the solutions.

In the following we solve \Eqs{dynamoeq}{emf_nonlocal} numerically
in terms of the mean magnetic
vector potential $\meanAA$, so $\meanBB=\nab\times\meanAA$.
We use here the {\sc Pencil Code}\footnote{%
\url{http://pencil-code.googlecode.com/}}, which is a high-order
public domain code (sixth order in space and third order in time) for
solving partial differential equations, including a range of different
mean-field equations.
The final set of equations is for vanishing mean flow
\EQ
{\partial\meanAA\over\partial t}=\meanEMF + \eta \frac{\partial^2 \meanAA}{\partial z^2},
\EN
\EQ
{\partial\meanEMF\over\partial t}=\alpha\meanBB-\etat\meanJJ
-{\meanEMF\over\tau}+\Mu{\partial^2\meanEMF\over\partial z^2},
\EN
of which only the $x$ and $y$ components are relevant.
We have introduced here the additional parameter $\Mu=\ell^2/\tau$
having the dimension of diffusivity.

For a simple dynamo with boundaries
we choose the $\alpha^2$ dynamo with a linear $\alpha$ profile,
\EQ
\alpha(z)=\alpha_0 z/L_z,
\label{alphaz}
\EN
with $0\leq z\leq L_z$, where $L_z = \pi/2 k_1$ is the size of the domain.
For the sake of simplicity we retain here the assumption of isotropy,
although it is strictly not tenable under inhomogeneous conditions.
Here, $k_1$ is the lowest wavenumber for a quarter-cosine wave
obeying the boundary conditions
\EQ
\meanA_{x,z}=\meanA_{y,z}=\meanemf_{x,z}=\meanemf_{y,z}=
0\quad\mbox{on $z=0$},
\EN
and
\EQ
\meanA_{x}=\meanA_{y}=\meanemf_{x}=\meanemf_{y}=
0\quad\mbox{on $z=L_z$}.
\EN
These conditions correspond to a perfect conductor condition on $z=L_z$
and select solutions $\meanBB$ antisymmetric about $z=0$.

We recall that the $\alpha^2$ dynamos with the linearly varying
$\alpha$ profile \eq{alphaz} are always oscillatory
with dynamo waves.
This was first noticed in direct numerical simulations (Mitra et al.\ 2010),
but was then also confirmed for mean-field models (Brandenburg et al.\ 2009)
and is consistent with the parametric survey of solutions given by
R\"udiger \& Hollerbach (2004).

\begin{figure}[t!]\begin{center}
\includegraphics[width=\columnwidth]{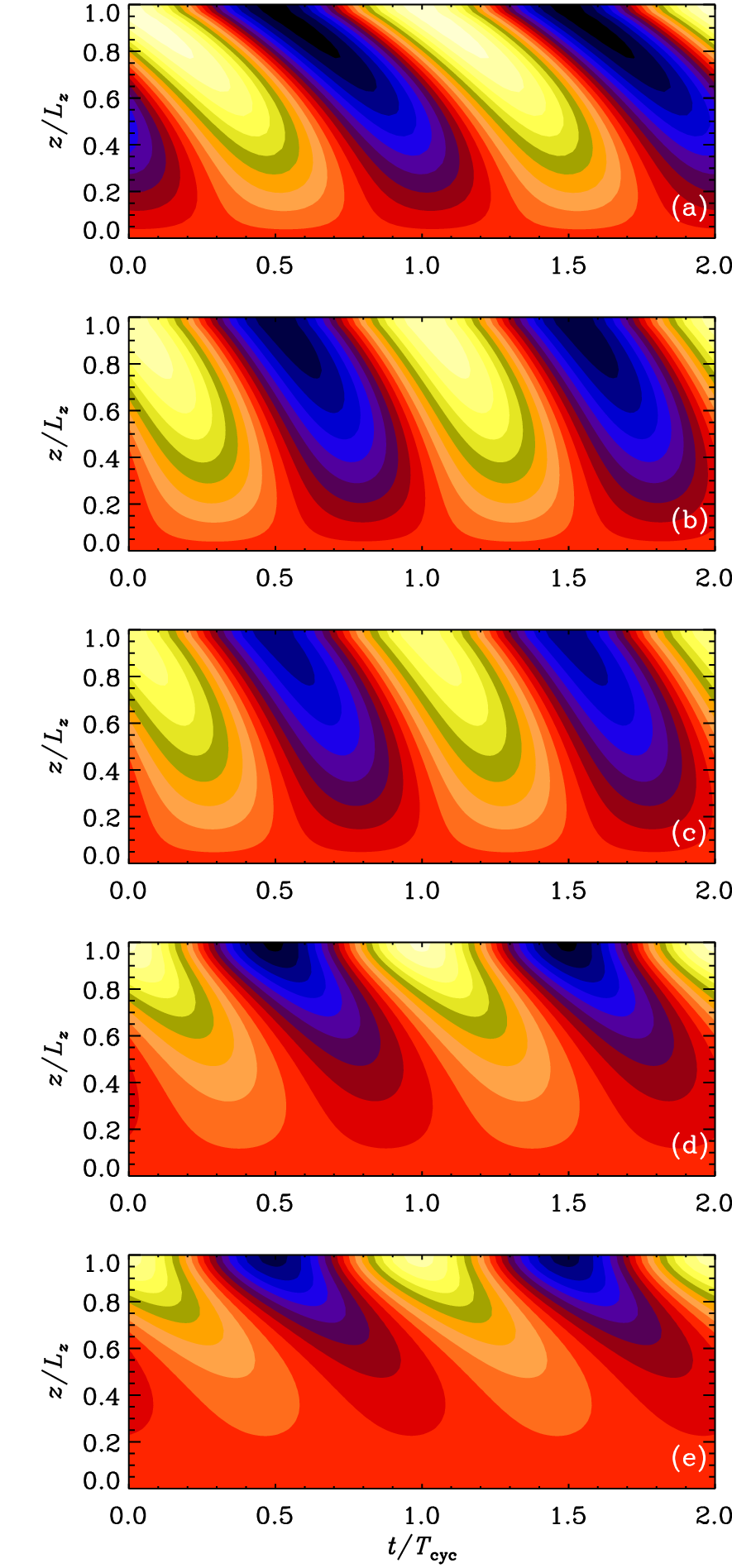}
\end{center}\caption[]{ 
Butterfly or $zt$ diagram of $\meanB_y$ for mean-field models
with different combinations of $\tau\etat k_1^2$ and $\Mu/\etat$.
(a): $\tau\etat k_1^2=\Mu/\etat=10^{-3}$;
(b) -- (d): $\tau\etat k_1^2=1$, $\Mu/\etat=10^{-3}$, 0.3, 1,
(e): $\tau\etat k_1^2=0.06$, $\Mu/\etat=3$.
$T_{\mathrm{cyc}}$ -- cycle period.
}\label{pbut_comp}\end{figure}

We have computed marginally excited dynamo solutions for different values
of $\tau\etat k_1^2$ and $\Mu/\etat$.
For comparison, for the value $\kf/k_1=2.2$ considered in \Sec{PeriodicDomain},
we have, using $\etat=\etatz$ and assuming $\tau/\tauto=\ell\kf=1$,
\EQ
\tau\etat k_1^2=(3\kf^2/k_1^2)^{-1}\approx0.06,
\quad
{\Mu\over\etat}={3\ell^2\kf^2\over\tau/\tauto}\approx3.
\EN
The critical values of the dynamo number $C_\alpha=\alpha/\etat k_1$
and the resulting normalized cycle
frequencies $\omega/\etat k_1^2$ are given in \Tab{Summary}.
For five particular cases, denoted by the
labels (a)--(e), the corresponding
butterfly diagrams are shown in \Fig{pbut_comp}.

\begin{table}[t!]\caption{
Dependence of $C_\alpha^{\rm crit}$ and
normalized cycle frequency $\omega/\etat k_1^2$
on $\tau\etat k_1^2$ and $\Mu/\etat$
for marginally excited solutions of $\alpha^2$ dynamos
with linear $\alpha$ profile \eq{alphaz}.
}\vspace{12pt}\centerline{\begin{tabular}{c|ccccccc}
Run &$\tau\etat k_1^2$ & $\Mu/\etat$ & $C_\alpha^{\rm crit}$ &
$\omega/\etat k_1^2$ \\[1mm]
\hline
\hline\\[-2.5mm]
(a) &0.001&0.001& 5.16 & 1.64 \\
    & 0.1 &0.001& 4.65 & 0.74 \\
(b) &  1  &0.001& 2.76 & 0.88 \\
    &  1  & 0.1 & 2.77 & 0.87 \\
(c) &  1  & 0.3 & 2.84 & 0.86 \\
    &  1  & 0.7 & 3.68 & 0.78 \\
(d) &  1  &   1 & 5.30 & 0.64 \\
(e) &0.06 &   3 & 8.12 & 0.58   
\label{Summary}\end{tabular}}\end{table}

Similar to the $\alpha$--shear dynamos discussed in \Sec{Nonlocality},
we find that the critical dynamo number
$C_\alpha^{\rm crit}$, is lowered in all cases with $\tau\neq0$;
see \Tab{Summary}.
Furthermore, and perhaps somewhat surprisingly, we find that, as
$\Mu/\etat$ is increased, the dynamo wave weakens significantly before
reaching the equator; see panels (b)--(e).
On the other hand, increasing $\tau\etat k_1^2$ from $10^{-3}$ to 1
does not affect the weakening of the dynamo wave near the equator,
but it rather enhances its speed.
Whether similar results also apply to $\alpha$--shear dynamos is
however not obvious.
Also, while the anticipated smoothing effect might explain the
weakening of the dynamo wave near the equator, it does not seem to
operate in the same way in the proximity of the boundary at $z=L_z$.
Instead, we see that the dynamo wave is now more nearly perpendicular
to that boundary compared with the case $\Mu\to0$.

\section{Conclusions}

The present work has established that the Fourier transform of the
integral kernel for the representation of the mean electromotive force
in the isotropic case is well approximated by
\EQ
{\hat{G}(k,\omega)} \propto \frac{1}{1-\ii\omega\tau+\ell^2 k^2},
\EN
which, in turn, can be captured by solving a partial differential
equation for the mean electromotive force with a first order time derivative
and a Laplacian that plays the role of a diffusion term.
Our work has illustrated the great ease with which
nonlocality in space and time can be implemented in a dynamo model.
Indeed, the
chosen,
 simplest possible kernel leads to a rather plausible
representation of the partial differential equation governing the
evolution of the electromotive force.
Furthermore, the application to spherical and other coordinate systems is
quite straightforward and already fully functional in the {\sc Pencil Code}.

It turns out that, while nonlocality normally hampers dynamo action,
it can actually make the dynamo more easily excitable provided
it is oscillatory.
This has here been shown analytically for standard dynamo
waves in the presence of shear, but it has also been found in the
case of an $\alpha^2$ dynamo in spherical geometry
where the oscillatory behavior is a consequence
of the spatial antisymmetry of $\alpha$ about the equatorial plane
(Mitra et al.\ 2010).

Another issue that has not been addressed here is the question of
nonlinearity.
Our present approach is easily extendable to the case where
$\alpha$ and $\etat$ are nonlinear functions of $\meanBB$,
as in the case of usual algebraic quenching.
Even the case of dynamic $\alpha$ quenching (Kleeorin \& Ruzmaikin 1982)
could easily be included.
Here, yet another differential equation is being solved,
namely one for a magnetic contribution to $\alpha$.
One might imagine that the effects of this additional equation
are already captured by the $\partial\meanEMF/\partial t$ equation.
However, it should be remembered that the dynamic $\alpha$ quenching
also contains effects of magnetic helicity fluxes and is capable of
reproducing the resistively slow saturation in the absence of such fluxes.

We regard the approach of solving a partial differential equation
for $\meanEMF$ as a natural one, which supersedes the usual
dynamo equations where $\tau\to0$ and $\ell\to0$ is assumed.
In many typical situations, neither of the two assumptions
are well satisfied.
We also recall that the approach of including the time derivative
of $\meanEMF$ addresses the problem of causality, i.e., the
propagation speed of disturbances of $\meanBB$
is automatically limited to the
value of the rms velocity of the turbulence, as demonstrated in 
Brandenburg et al.\ (2004).
Furthermore, the presence of the diffusion operator in the
evolution equation for $\meanEMF$ is natural and advantageous
because it ensures numerical stability and, more importantly,
it prevents, in a physical way, the emergence of artificially sharp
structures on scales comparable to or below that of the turbulence.
Is should be noted, however, that, while the time derivative
of $\meanEMF$
emerges as a natural consequence from the $\tau$ approach
(Blackman \& Field 2002, 2003), there does not seem to be
a likewise natural motivation for the presence of the diffusion term
in the equation for $\meanEMF$.

\acknowledgements
We acknowledge the NORDITA dynamo program of 2011 for
providing a stimulating scientific atmosphere.
The computations have been carried out on the
National Supercomputer Centre in Link\"oping and the Center for
Parallel Computers at the Royal Institute of Technology in Sweden.
This work was supported in part by the European Research Council
under the AstroDyn Research Project 227952.


\begin{appendix}

\section{Dispersion relations for nonlocal dynamos}

\subsection{$\alpha^2$ dynamos}
\label{Disper}

We begin by writing the governing equations \eqss{dynamoeq}{emf_nonlocal}
in component form
for homogeneous turbulence, i.e. constant mean-field coefficients, hence
\EQ
{\partial\meanB_x\over\partial t}
=-{\partial\meanemf_y\over\partial z}+\eta{\partial^2\meanB_x\over\partial z^2},
\EN
\EQ
{\partial\meanB_y\over\partial t}
=+{\partial\meanemf_x\over\partial z}+\eta{\partial^2\meanB_y\over\partial z^2},
\EN
\EQ
\meanemf_x
+\tau{\partial\meanemf_x\over\partial t}
-\ell^2{\partial^2\meanemf_x\over\partial z^2}
=\alpha\meanB_x+\etat{\partial\meanB_y\over\partial z},
\EN
\EQ
\meanemf_y
+\tau{\partial\meanemf_y\over\partial t}
-\ell^2{\partial^2\meanemf_y\over\partial z^2}
=\alpha\meanB_y-\etat{\partial\meanB_x\over\partial z}.
\EN
The dispersion relation is easily obtained by
employing the ansatz \eq{Bansatz} with $\omega=0$ in these equations and 
 writing them
in matrix form, $\MMMM\qq=\bm{0}$, where
$\qq=(\meanB_x, \meanB_y, \meanemf_x, \meanemf_y)^T$
is the state vector and
\[
\MMMM=\pmatrix{
\lambda+\eta k^2 & 0 & 0 & +\ii k\cr
0 & \lambda+\eta k^2 & -\ii k & 0 \cr
-\alpha & -\ii k\etat & \!\!1+\lambda\tau+\ell^2 k^2\!\! & 0 \cr
+\ii k\etat & -\alpha & 0 & \!\!1+\lambda\tau+\ell^2 k^2 \!}
\]
is the matrix $\MMMM$ for the $\alpha^2$ dynamo.
Nontrivial solutions have vanishing determinant,
which yields
\EQ
\left[\left(\lambda+\eta k^2\right)
\left(1+\lambda\tau+\ell^2 k^2\right)+\etat k^2\right]^2
=\alpha^2k^2.
\EN
Taking the square root leads to the implicit solution \eq{lama2}.

\subsection{$\alpha$--shear dynamos}
\label{Disper2}

In the case of a pure $\alpha$--shear dynamo with a mean flow
of the form $\meanUU=(0,Sx,0)$, and neglecting
the terms $\alpha\meanB_x$, we have
\EQ
{\partial\meanB_x\over\partial t}=
-{\partial\meanemf_y\over\partial z}+\eta{\partial^2\meanB_x\over\partial z^2},
\EN
\EQ
{\partial\meanB_y\over\partial t}=S\meanB_x
+{\partial\meanemf_x\over\partial z}+\eta{\partial^2\meanB_y\over\partial z^2},
\EN
\EQ
\meanemf_x
+\tau{\partial\meanemf_x\over\partial t}
-\ell^2{\partial^2\meanemf_x\over\partial z^2}
=+\etat{\partial\meanB_y\over\partial z},
\EN
\EQ
\meanemf_y
+\tau{\partial\meanemf_y\over\partial t}
-\ell^2{\partial^2\meanemf_y\over\partial z^2}
=\alpha\meanB_y-\etat{\partial\meanB_x\over\partial z}.
\EN
In the marginally excited oscillatory case, the matrix $\MMMM$ is
for $\eta=0$
\EQ
\pmatrix{
-\ii\omega & 0 & 0 & \ii k\cr
-S & -\ii\omega & -\ii k & 0 \cr
0 & -\ii k\etat & \Iota-\ii\omega\tau & 0 \cr
\ii k\etat & -\alpha & 0 & \Iota-\ii\omega\tau }
\pmatrix{\meanB_x\cr\meanB_y\cr\meanemf_x\cr\meanemf_y}=0,
\EN
$\Iota=1+\ell^2 k^2$.
The dispersion relation becomes
\EQ
\left[-\ii\omega(\Iota-\ii\omega\tau)+\etat k^2\right]^2
+\ii k\alpha S(\Iota-\ii\omega\tau)=0.
\EN
Solving separately for real and imaginary parts, we obtain
\EQ
-\omega^2\Iota^2+(-\omega^2\tau+\etat k^2)^2
+\etat^2 k^4+k\alpha S\omega\tau=0
\EN
and
\EQ
2\omega(\omega^2\tau-\etat k^2)+k\alpha S=0.
\EN
Eliminating $k\alpha S$ yields then \Eqs{omaS}{Dcrit}.

\end{appendix}
\end{document}